\documentclass[aps,pra,twocolumn,amsmath,amssymb,nofootinbib,showpacs,superscriptaddress]{revtex4}
\usepackage[english]{babel}
\usepackage{latexsym}
\usepackage{graphics}
\usepackage{graphicx}
\usepackage{epsfig}
\usepackage{bm}
\usepackage{amsmath}
\usepackage{amssymb}
\usepackage{amsthm}
\usepackage{dcolumn}
\usepackage{bm}
\usepackage{soul}
\usepackage{subfig}
\usepackage[dvipsnames]{xcolor}
\usepackage{float}
\usepackage{tikz} 
\usetikzlibrary{calc}
\usepackage{pgfplots}
\pgfplotsset{compat=newest, width=8 cm}%версия пакета построения графиков, ширина графиков
\usetikzlibrary{arrows.meta}
\usepgfplotslibrary{fillbetween}
\usepgfplotslibrary{polar}
\usepgfplotslibrary{fillbetween}
\usepgfplotslibrary{polar}
\usepackage{pgfplotstable}
\usepackage[bookmarks=false]{hyperref}
\hypersetup{
    colorlinks,
    linkcolor={red!90!black},
    citecolor={blue!90!black},
    urlcolor={blue!80!black}
}
\usepackage[sort&compress]{natbib}
\graphicspath{{Images/}}
\usepackage{comment}
\usepackage{epstopdf}
\usepackage{cancel}
\usepackage{cleveref}
\usepackage{enumerate}
\usepackage{braket}
\usepackage{lipsum}  
\usepackage[hybrid]{markdown}
\usepackage{appendix}
\theoremstyle{remark}

\begin{document}

\title{Photon pairs, squeezed light and the quantum wave mixing effect in a cascaded qubit system}
\author{R. D. Ivanovskikh}
\email{roman\_ivskh@mail.ru} 
\affiliation{Dukhov Research Institute of Automatics (VNIIA), Moscow 127055, Russia}
\author{W. V. Pogosov}
\affiliation{Dukhov Research Institute of Automatics (VNIIA), Moscow 127055, Russia}
\affiliation{Moscow Institute of Physics and Technology, Dolgoprudny, 141700, Russia}
\affiliation{Institute for Theoretical and Applied Electrodynamics, Russian Academy of Sciences, Moscow, 125412, Russia}
\author{A. A. Elistratov}
\affiliation{Dukhov Research Institute of Automatics (VNIIA), Moscow 127055, Russia}
\author{S.V.Remizov}
\affiliation{Dukhov Research Institute of Automatics (VNIIA), Moscow 127055, Russia}
\affiliation{Kotel’nikov Institute of Radio Engineering and Electronics, Russian Academy of Sciences, Moscow 125009, Russia}
\affiliation{HSE University, Moscow 109028, Russia}
\author{A. Yu. Dmitriev}
\affiliation{Moscow Institute of Physics and Technology, Dolgoprudny, 141700, Russia}
\author{T. R. Sabirov}
\affiliation{Skolkovo Institute of Science and Technology, Nobel St. 3, 143026 Moscow, Russia}
\affiliation{Moscow Institute of Physics and Technology, Dolgoprudny, 141700, Russia}
\author{A. V. Vasenin} 
\affiliation{Moscow Institute of Physics and Technology, Dolgoprudny, 141700, Russia}
\author{S.A. Gunin} 
\affiliation{Moscow Institute of Physics and Technology, Dolgoprudny, 141700, Russia}
\author{O. V. Astafiev}
\affiliation{Skolkovo Institute of Science and Technology, Nobel St. 3, 143026 Moscow, Russia}
\affiliation{Kotel’nikov Institute of Radio Engineering and Electronics, Russian Academy of Sciences, Moscow 125009, Russia}
\affiliation{Moscow Institute of Physics and Technology, Dolgoprudny, 141700, Russia}

 %       \lipsum[3]

\begin{abstract}
We develop a theoretical description of quantum wave mixing (QWM) in a cascaded
waveguide-QED system of two superconducting qubits, where the probe is driven by
an external coherent tone and by the resonance fluorescence of a strongly driven
source qubit. Starting from the field correlation functions of the source
emission, we derive an effective master-equation treatment for the probe and
identify the regime in which the incident fluorescence is characterized by anomalous correlations. When the coherent Rayleigh component
of the source spectrum is suppressed, the probe equations of motion become
equivalent to those for a qubit driven by a coherent tone and broadband squeezed
light, although the ligt emitted by the source is non-Gaussian and antibunched. This equivalence implies a selection rule 
for the peaks of the QWM
spectrum, with a strong suppression of sidebands associated with processes
involving an odd number of photons taken from the source field. Numerical
simulations of the full cascaded two-qubit model for different ratios of
radiative decay rates unambiguously
confirm the participation of correlated photon pairs in QWM processes. The current research illustrates that the analysis of peak amplitudes can be used to probe photon statistics in the incident nonclassical field.
\end{abstract}

\maketitle

\section{Introduction}
Superconducting qubits strongly coupled to a one-dimensional waveguide
\cite{Astafiev2010resonance, Delsing2013microwave1Dspace} provide a controllable
platform for studying nonlinear quantum optics in the microwave domain. In this
architecture, a single qubit acts as a highly nonlinear scatterer for
propagating fields, enabling the observation of a broad range of effects,
including resonance fluorescence \cite{Abdumalikov2011dynamics, Vasenin2024evolution},
superradiance \cite{mlynek2014observation}, and the formation of dark states
\cite{mirhosseini2019cavity, Yin2023generation}. Beyond their intrinsic interest,
such phenomena offer experimentally accessible signatures of multiphoton
scattering processes at the level of a single artificial atom.

Quantum wave mixing (QWM) is a representative example of such single-qubit
nonlinear response. When a two-level system is driven by two waves with
different carrier frequencies, the scattered radiation contains a set of
phase-coherent sidebands at frequencies given by integer linear combinations of
the drives. These components can be understood as elastic multiphoton
scattering events: the total energy is conserved in each event, while photons
are exchanged between the incident modes and the scattered field. In contrast
to conventional wave mixing in macroscopic nonlinear media, QWM is governed by
the quantum dynamics of a single emitter and can therefore become sensitive to
the quantum state of the incoming radiation.

The QWM spectrum of a superconducting qubit was first observed for trains of
pulses \cite{Dmitriev_2017} and for continuous coherent tones \cite{Dmitriev2019}.
In these experiments the sidebands appear as narrow peaks whose weights encode
the relative importance of different multiphoton scattering pathways. A key
extension, conjected already in Ref.~\cite{Dmitriev2019}, is that the amplitudes
of these peaks need not be universal: they can change when one of the incident
fields is nonclassical, thus providing access to photon statistics via a purely
spectral measurement. This idea was analyzed in detail in Ref.~\cite{effects}.

In particular, Ref.~\cite{effects} addressed QWM with one coherent drive and one
nonclassical field and demonstrated that the resulting spectra can exhibit
selection rules absent for two coherent waves. Two representative examples were
considered. First, broadband squeezed light, whose properties and generation
have been extensively studied theoretically and experimentally
\cite{Walls,Gardiner_DPA,Schnabel}, which in circuit QED can be
generated using a degenerate parametric amplifier (DPA)
\cite{PhysRevX.6.031004, Q_Noise}, modifies the multiphoton scattering channels
in a way that can suppress specific families of sidebands. Second, a
periodically prepared superposition of Fock states $\ket{0}$ and $\ket{1}$,
realizable by using another qubit as a single-photon source
\cite{single-photon-source}, also leads to a qualitatively altered mixing
spectrum. Related wave-mixing effects in multilevel emitters were studied, for
example, for a three-level atom in Ref.~\cite{PhysRevA.98.041801}.

A complementary route to nonclassical driving is provided by cascaded
waveguide-QED systems, where radiation is emitted by the source qubit to the probe qubit. In such a geometry, the probe can scatter the
source emission together with an externally applied coherent signal, producing
a QWM spectrum that reflects the statistics of the field generated by the
source. A
microwave implementation of a two-atom cascade aimed at probing such effects
was proposed in Ref.~\cite{Vasenin_aip}. This regime was investigated experimentally and numerically in
Ref.~\cite{PhysRevA.111.043715} for different ratios of the qubit radiative decay
rates. In that setting, the nonclassical component is provided by the resonance
fluorescence of the driven source qubit.

Resonance fluorescence of a coherently driven two-level system provides one of
the simplest microscopic mechanisms by which a classical driving field is
converted into nonclassical radiation. In addition to antibunching and
sub-Poissonian photon statistics, the emitted field may exhibit reduced
quadrature fluctuations. Squeezing in resonance fluorescence was predicted by
Walls and Zoller~\cite{WallsZoller1981}, while its frequency-resolved form was
analyzed by Collett, Walls, and Zoller in terms of the spectrum of
squeezing~\cite{CollettWallsZoller1984}. Later theoretical studies clarified
how the squeezing depends on the excitation regime: weak resonant driving leads
to narrow-band squeezing close to the coherent scattering frequency, whereas
detuned and stronger excitation can give rise to two-mode squeezing associated
with the sidebands of the Mollow spectrum~\cite{ZhouSwain1999}. These
predictions have also been supported experimentally in solid-state two-level
systems, where quadrature squeezing of resonance-fluorescence
photons~\cite{Schulte2015} and intensity squeezing in a
resonance-fluorescence source~\cite{Wang2020} have been observed.

The existence of quadrature squeezing in resonance fluorescence does not by
itself determine the nonlinear response of a downstream emitter. The complete
field radiated by a two-level system is antibunched and strongly non-Gaussian,
and is therefore not globally equivalent to the broadband squeezed field
considered in Ref.~\cite{effects}. Frequency selection is essential: a probe
with $\gamma_{\rm{pr}}\ll\gamma_{\rm{s}}$ samples a narrow spectral window
around the central Mollow component, where frequency-resolved bunching and a
nonvanishing anomalous correlator can coexist with antibunching of the
unfiltered fluorescence \cite{Nienhuis_1,Nienhuis_2, Arnoldus_1984}. The equivalence derived
below is thus an operational, bandwidth-dependent equivalence of the reduced
probe dynamics, rather than an identity between the quantum state emitted by
the source qubit and that produced by a DPA. In the cascaded architecture, the same two-qubit waveguide circuit generates
the nonclassical input, spectrally selects it, and converts its correlations
into QWM sidebands. In the broadband regime, the driven source qubit can thus
serve as an effective squeezed-field source, avoiding the need for a separate
DPA and thereby simplifying the experimental implementation.

The numerical and experimental results of Ref.~\cite{PhysRevA.111.043715} show
that for moderate source pumping the probe QWM spectrum is dominated by
multiphoton processes involving at most one photon originating from the source
field. However, as shown in the present paper, the picture changes when the source is driven so strongly that
the coherent part of the Rayleigh peak is suppressed. In this regime the probe
spectrum acquires a pronounced even/odd asymmetry: sidebands associated with
processes involving an odd number of photons from the source become strongly
suppressed. Establishing a transparent theoretical interpretation of this
selection rule and identifying its connection to anomalous correlations
of the source emission is the main motivation for the present work. 

To describe this strong-driving behaviour, we calculate the normal and anomalous
correlation functions of the source fluorescence and derive an effective master
equation and closed equations of motion for the probe. In the broadband-source limit, the probe dynamics is described by
effective normal and anomalous parameters $N$ and $M$ that are generated
microscopically by the source qubit and sampled within the probe bandwidth,
reproducing the equations of Ref.~\cite{effects}. The full cascaded model then
establishes the limits of this reduction: the QWM selection rule survives even
for $\gamma_{\rm{s}}\sim\gamma_{\rm{pr}}$ because it originates from the sign
symmetry of the central dressed-state component, whereas the allowed sidebands
exhibit a nonmonotonic finite-bandwidth enhancement resulting from the
competition between pair-correlated driving and source-induced probe
broadening.

The work is organized as follows. In Sec.~\ref{sec::Strong drive} we describe a
qubit under a strong coherent drive and evaluate its normal and anomalous
correlators. In Sec.~\ref{sec::Master equation} we derive a master equation for
the probe qubit driven by a field characterized by these source correlators,
and obtain closed equations of motion for the probe averages
$\braket{\sigma_{-}}$ and $\braket{\sigma_{z}}$. In Sec.~\ref{sec::Discussion} we
discuss the implications for the QWM spectrum and compare the analytical
description with numerical simulations.

\section{Qubit under strong drive}\label{sec::Strong drive}

We start from the Lindblad master equation for the driven source qubit,
\begin{equation}\label{Linblad}
\dfrac{d\rho}{dt}=-iL\rho,
\end{equation}
where
\begin{equation}\label {Linbladian}
-iL\rho=-i[H,\rho]/\hbar-\Gamma\rho,
\end{equation}
\begin{equation}\label{Hamiltonian}
H=-\hbar(\Delta\sigma_{z}+\Omega_{\rm{s}}\sigma_{x}),
\end {equation}
and $\Omega_{\rm{s}}$ is the Rabi frequency of the coherent drive applied to the
source. The Hamiltonian is written in a rotating frame defined by the source
drive frequency $\omega_{\rm{s}}$, and $\Delta=\omega_{\rm{s}}-\omega_{0}$ is the
detuning from the qubit transition frequency $\omega_{0}$.

The dissipator $\Gamma$ in Eq.~(\ref{Linbladian}) is taken in the standard
radiative form,
\begin{equation}\label {decay operator}
\Gamma\rho=\frac{1}{2}\gamma_{\rm{s}}(\sigma^{+}\sigma^{-}\rho+\rho\sigma^{+}\sigma^{-}-2\sigma^{-}\rho\sigma^{+}),
\end {equation}
where $\gamma_{\rm{s}}$ is the decay rate of the source qubit.

Following \cite{Nienhuis_1}, we introduce the dressed-state basis which
diagonalizes the Hamiltonian,
\begin{equation}
\begin{split}
&\ket{1}=c\ket{g}-s\ket{e},\\ &\ket{2}=s\ket{g}+c\ket{e},
\end{split}
\end{equation}
with
\begin{equation}
c=\sqrt{\dfrac{\Omega^{\prime}+\Delta}{2\Omega^{\prime}}},\ \ \ s=\sqrt{\dfrac{\Omega^{\prime}-\Delta}{2\Omega^{\prime}}},
\end{equation}
where $\Omega^{\prime}=\sqrt{\Omega^2+\Delta^2}$. The corresponding eigenvalues
are
\[
E_{1,2}=\pm\hbar\Omega^{\prime}.
\]

In this basis the raising and lowering operators can be decomposed as
\begin{equation}
\sigma^{\pm}=\sigma_{F}^{\pm}+\sigma_{T}^{\pm}+\sigma_{R}^{\pm}.
\end{equation}
Here
\begin{equation}\label{sec2::Pauli new variables}
\begin{split}
&\sigma_{F}^{-}=c^{2}\ket{1}\bra{2},\ \ \sigma_{T}^{-}=-s^{2}\ket{2}\bra{1},\\ &\sigma_{R}^{-}=cs(\ket{2}\bra{2}-\ket{1}\bra{1}).
\end{split}
\end{equation}
The three components correspond to the central ($R$) and sideband ($F$ and $T$)
contributions of resonance fluorescence in the dressed-state picture.

In the limit $\gamma\ll\Omega^{\prime}$, where the fluorescence triplet peaks are well
separated, the dissipator (\ref{decay operator}) can be written in the secular
form,
\begin{equation}\label{decay operator simplified}
\Gamma_{d}\rho=\frac{1}{2}\gamma_{\rm{s}}\sum\limits_{\alpha=F,R,T}[\sigma_{\alpha}^{+}\sigma_{\alpha}^{-}\rho+\rho\sigma_{\alpha}^{+}\sigma_{\alpha}^{-}-2\sigma_{\alpha}^{-}\rho\sigma_{\alpha}^{+}].
\end{equation}

Equations (\ref{Linblad})--(\ref{Hamiltonian}) together with
Eq.~(\ref{decay operator simplified}) yield the following equations for the
density-matrix elements in the dressed basis:
\begin{equation}\label{sec2::density comp equations}
\begin {split}
&\dot\rho_{11}=-\dot\rho_{22}=\gamma_{\rm{s}}(c^{4}\rho_{22}-s^{4}\rho_{11}),\\ &\dot\rho_{12}=-(\gamma^{\prime}-i\Omega^{\prime})\rho_{12}, \ \  
\dot\rho_{21}=-(\gamma^{\prime}+i\Omega^{\prime})\rho_{21},
\end{split}
\end{equation}
where $\gamma^{\prime}=\frac{\gamma_{\rm{s}}}{2}(1+2c^{2}s^{2})$.

The stationary solution of Eqs.~(\ref{sec2::density comp equations}) is
\begin{equation}\label{sec2::eq solution}
\bar\rho_{11}=\dfrac{c^{4}}{c^{4}+s^{4}},\ \ \bar\rho_{22}=\dfrac{s^{4}}{c^{4}+s^{4}},\ \ \bar\rho_{12}=\bar\rho_{21}=0.
\end{equation}

The normal correlators
$I_{\alpha}(t+\tau,t)=\gamma_{\rm{s}}\braket{\sigma_{\alpha}^{+}(t+\tau)\sigma_{\alpha}^{-}(t)}$
form the fluorescence triplet \cite{Apanasevich}, \cite{Nienhuis_1}, \cite{mollow_pioneer}
and determine the spectrum emitted by a strongly driven two-level system:
\begin{widetext}
\begin{equation}\label{sec2::I_R solution}
I_{R}(\omega,\omega^{\prime})=\left(I_{Ri}\dfrac{1}{i(\omega-\omega_{2})-\gamma_{0}}+I_{Rc}2\pi\delta(\omega-\omega_{2})\right)\delta(\omega-\omega^{\prime}),  
\end{equation}
\begin{equation}\label{sec2::I_F solution}
I_{F}(\omega, \omega^{\prime})=I_{F}\dfrac{1}{i(\omega-(\omega_{\rm{s}}-\Omega^{\prime}))-\gamma^{\prime}}\delta(\omega-\omega^{\prime}),
\end{equation}
\begin{equation}\label{sec2::I_T solution}
I_{T}(\omega, \omega^{\prime})=I_{T}\dfrac{1}{i(\omega-(\omega_{\rm{s}}+\Omega^{\prime}))-\gamma^{\prime}}\delta(\omega-\omega^{\prime}),
\end{equation}
\begin{equation}\label{sec2::coh and incoh}
I_{Rc}=\gamma_{\rm{s}}{c}^{2}s^{2}\left(\dfrac{c^{4}-s^{4}}{c^{4}+s^{4}}\right)^2,\ \ I_{Ri}=\gamma_{\rm{s}}{c}^{2}s^{2}\dfrac{4c^{4}s^{4}}{({c^{4}+s^{4}})^2}.
\end{equation}
\begin{equation}\label{sec2::FT bands}
I_{F}=I_{T}=\gamma_{\rm{s}}\dfrac{c^{4}s^{4}}{c^{4}+s^{4}},
\end{equation}
where $\gamma_{0}=\gamma_{\rm{s}}(c^{4}+s^{4})$,  $\gamma^{\prime}=\frac{\gamma_{\rm{s}}}{2}(1+2c^{2}s^{2})$. Indices R,T,F determine the peak of the fluorescence triplet: R is the central peak, F and T are the sidebands. The central or the Rayleigh component of the fluorescence triplet contains fluorescent emission $I_{Ri}$ and a coherent fraction $I_{Rc}$ corresponding to elastic scattering. The latter disappears when $\Omega_{\rm{s}}\gg\Delta$ and $c^2\approx s^{2}$.  
\end{widetext}

We now evaluate the anomalous (phase-sensitive) two-time correlators
$F_{\alpha\beta}(t+\tau,t)=\gamma\braket{\sigma_{\alpha}^{+}(t+\tau)\sigma_{\beta}^{+}(t)}$
using the quantum regression theorem, following \cite{Nienhuis_1}.
From Eqs.~(\ref{sec2::density comp equations}) one obtains an equation of motion
for $\braket{\sigma_{R}^{+}}$:
\begin{equation}
\frac{\partial}{\partial{t}}\braket{\sigma_{R}^{+}}=-\gamma{c}{s}(c^{2}-s^{2})-\gamma(1-2c^{2}s^{2})\braket{\sigma_{R}^{+}}.
\end{equation}

Applying the quantum regression theorem gives the corresponding evolution
equation for the two-time expectation value,
\begin{equation}\label{sec2::RR equation}
\begin{split}
\frac{\partial}{\partial\tau}\braket{\sigma_{R}^{+}(t+\tau)\sigma_{R}^{+}(t)}&=-\gamma{c}{s}(c^{2}-s^{2})\braket{\sigma_{R}^{+}(t)}-\\&\gamma(1-2c^{2}s^{2})\braket{\sigma_{R}^{+}(t+\tau)\sigma_{R}^{+}(t)}
\end{split}
\end{equation}
with the initial condition following from (\ref{sec2::eq solution}):
\[
\braket{\sigma^{+}_{R}(t)\sigma^{+}_{R}(t)}=c^{2}s^{2}.
\]

Solving Eq.~(\ref{sec2::RR equation}) with this initial condition yields
\[
F_{RR}(t,\tau)=I_{Rc}+I_{Ri}e^{-\gamma(1-2c^{2}s^{2})\tau},
\]
or, after transforming from the rotating frame to the laboratory one
$\braket{\sigma^{+}_{R}(t)}\rightarrow\braket{\sigma^{+}_{R}(t)}e^{-i\omega_{\rm{s}}t}$
and using $\gamma_{\rm{s}}(1-2c^{2}s^{2})=\gamma_{0}$,
\begin{equation}\label{sec2::RR anomal in time}
F_{RR}(t+\tau,t)=\left(I_{Rc}+I_{Ri}e^{-\gamma_{0}\tau}\right)e^{-i\omega_{\rm{s}}(t+\tau)}e^{-i\omega_{\rm{s}}t},
\end{equation}
where $I_{Rc}$ and $I_{Ri}$ are defined in
(\ref{sec2::coh and incoh}-\ref{sec2::FT bands}). Taking the Fourier transform
of (\ref{sec2::RR anomal in time}),
\[
F_{RR}(\omega,\omega^{\prime})=\gamma_{\rm{s}}\int{d}t\int{d}t^{\prime}\braket{\sigma_{R}^{+}(t^{\prime})\sigma_{R}^{+}(t)}e^{i\omega^{\prime}{t}^{\prime}+i\omega{t}},
\]
we obtain
\begin{widetext}
\begin{equation}\label{sec2::RR anomal}
F_{RR}(\omega,\omega^{\prime})=2\pi\delta(\omega+\omega^{\prime}-2\omega_{\rm{s}})\left(I_{Rc}2\pi\delta(\omega-\omega^{\prime})+I_{Ri}\dfrac{1}{i(\omega^{\prime}-\omega_{\rm{s}})-\gamma_{0}}  \right),
\end{equation}
\end{widetext}
where $\gamma_{0}=\gamma_{\rm{s}}(c^{4}+s^{4})$. As for the normal correlators,
$F_{RR}$ contains both coherent and incoherent parts; the coherent contribution
vanishes in the strong-drive limit.

Similarly, one finds
\begin{widetext}
\begin{equation}\label{sec2::FT anomal}
    F_{FT}(\omega,\omega^{\prime})=F_{FT}2\pi\delta(\omega+\omega^{\prime}-2\omega_{\rm{s}})\frac{1}{i(\omega^{\prime}-(\omega_{\rm{s}}+\Omega^{\prime}))-\gamma^{\prime}}, 
\end{equation}
where 
\[
F_{FT}=\gamma_{\rm{s}}\braket{\sigma_{F}^{+}(t)\sigma_{T}^{+}(t)}=\gamma_{\rm{s}}(-s^{2}c^{2})\dfrac{s^{4}}{c^{4}+s^{4}}
\]
is the simultaneous anomal correlator. 
\end{widetext}

The presence of anomalous correlators indicates phase-sensitive pair
correlations in the fluorescence field, with the total energy fixed by the
drive frequency through the constraint $\omega+\omega' = 2\omega_{\rm{s}}$.
Anomalous correlators such as $F_{RF}$, $F_{RT}$, $F_{FF}$, etc., are equal to
zero; physically, these correlations are absent because the corresponding
processes do not satisfy the energy constraint.

In the next sections we show how these pair correlations, together with the
suppression of the coherent Rayleigh component in the strong-drive regime,
determine the structure of the QWM spectrum in the cascaded setup.

\section{Master equation for the probe qubit}\label{sec::Master equation}

We now consider the probe qubit interacting with the field produced by the
source, focusing on the frequency range of the fluorescence triplet central component.
The Hamiltonian of a qubit coupled to an electromagnetic field in a waveguide is
given by
\begin{equation}\label{H general}
H=H_{\rm{B}}+H_{\rm{sys}}+H_{\rm{int}}, 
\end{equation}
where the Hamiltonian of the photon reservoir is
\begin{equation}\label{H bath}
H_{B}=\sum_{p}\hbar\omega_{p}a_{p}^{\dagger}a_{p},
\end {equation}
the Hamiltonian of the probe qubit under the coherent drive with the frequency
$\omega_{\rm{pr}}$ and the amplitude $\Omega_{\rm{pr}}$:
\begin{equation}\label{H qbit}
\begin{split}
H_{\rm{sys}}=&\dfrac{1}{2}\hbar\omega_{0}\sigma_{z}+\\&\left(\Omega_{\rm{pr}}e^{-i\omega_{\rm{pr}}t}\sigma_{+}^{\rm{pr}}+\text{h.c.} \right)
\end{split}
\end{equation}
where $\omega_{0}$ is the transition frequency of the probe qubit (for
simplicity is taken equal to the transition frequency of the source qubit).
Here and below, the Pauli operators refer to the probe qubit. The interaction
Hamiltonian reads
\begin{equation}\label{H int}
H_{\rm{int}}=\hbar\sum_{i}\sigma_{i}\Gamma_{i},
\end{equation}
where $i=\pm$, and
\[
\Gamma_{-}=\sum_{p}\kappa_{p}a^{\dagger}_{p},\ \ \Gamma_{+}=\Gamma_{-}^{*}.
\]

For this system the master equation can be written in the form \cite{Carmichael}
\begin{widetext}
\begin{equation}\label{sec3::master eq general}
\begin{split}
\dot\rho(t)=-\frac{i}{\hbar}\left[H_{\rm{sys}},\rho\right]-\sum_{i,j=\pm}\int\limits_{0}^{t}dt^{\prime}&\left[\sigma_{i}\sigma_{j}\rho(t^{\prime})-\sigma_{j}\rho(t^{\prime})\sigma_{i}\right]\braket{\Gamma_{i} (t)\Gamma_{j}(t^{\prime})}+\\&\left[\rho(t^{\prime})\sigma_{j}\sigma_{i}-\sigma_{i}\rho(t^{\prime})\sigma_{j}\right]\braket{\Gamma_{j}(t^{\prime})\Gamma_{i}(t)},
\end{split}
\end{equation}
where the time dependence of the reservoir operators is determined by the interaction representation. 
\end{widetext}

The term with $i=j=-$ involves the anomalous reservoir correlator and
corresponds to the ``squeezed'' contribution,
\begin{widetext}
\begin{equation}\label{squeezed part}
\dot\rho\propto\int\limits_{0}^{t}dt^{\prime}\sigma_{-}\rho(t^{\prime})\sigma_{-}\sum_{p,p^{\prime}}\kappa_{p}\kappa_{p^{\prime}}\braket{a_{p}^{\dagger}(t)a_{p^{\prime}}^{\dagger}(t^{\prime})}+\left\{p\leftrightarrow{p}^{\prime},t\leftrightarrow{t^{\prime}}\right\}.
\end{equation}
\end{widetext}

To evaluate the reservoir correlators, we express the outgoing field of the
source qubit via the input--output relation,
\begin{equation}\label{sec3:in-out}
a_{\rm{out}}(\omega_p)=a_{\rm{in}}(\omega_p)+\sqrt{\gamma_{\rm{s}}}\sigma_{R}(\omega_{p}),
\end{equation}
and assume that the ``input'' field $a_{\rm{in}}$ incident on the source is
vacuum (the coherent pump on the source is already included in the source
Hamiltonian),
\[
\braket{a_{\rm{in}}(t) a^{\dagger}_{\rm{in}}(t^{\prime})}=\delta(t-t^{\prime}), \ \ \ \ \ \ \braket{a^{\dagger}_{\rm{in}}(t), a_{\rm{in}}(t^{\prime})}=0.
\]

\begin{figure}[!t]
    \includegraphics[width=7cm, height=3cm]{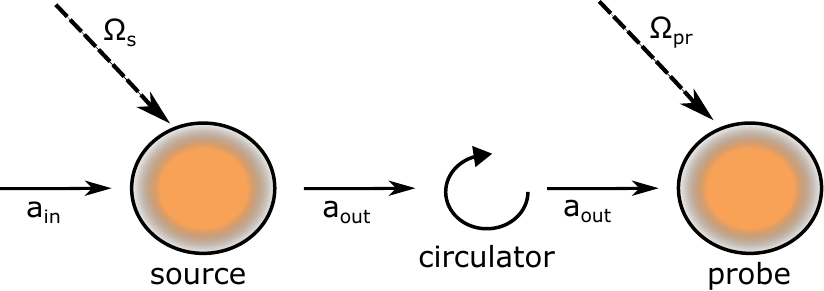} 
    \caption{\raggedright Cascaded source--probe system. The circulator enforces unidirectional coupling: radiation emitted by the source drives the probe, while back-action from the probe to the source is suppressed. The input field $a_{\rm{in}}$ incident on the source is assumed to be vacuum. The coherent pump on the source is not included in $a_{\rm{in}}$.}\label{fig:cascaded}
\end{figure}

Because the system is cascaded, the probe does not act back on the source; the
probe dynamics is driven by the field emitted by the source, as illustrated in
Fig.~\ref{fig:cascaded}.

Correlators of the form $\braket{a_{\rm{in}}^{\dagger}(\omega_p)\sigma_{R}(\omega_{p^{\prime}})}$
vanish. A convenient way to see this is to use the quantum Langevin formulation:
the Lindblad generator (\ref{Linblad}) with the dissipator in the form
(\ref{decay operator simplified}) is equivalent \cite{Q_Noise} to Langevin
equations for the operators $\sigma_{R,F,T}$. For the particular case
$c^{2}=s^{2}=1/2$ these equations read
\[
\dot{\sigma_{R}}=-\dfrac{\gamma}{2}\sigma_{R}+\sqrt{\gamma}(a^{\dagger}_{\rm{in}}\sigma_{T}+\sigma_{T}^{+}a_{\rm{in}})-\sqrt{\gamma}(\sigma_{F}^{+}a_{\rm{in}}+a_{\rm{in}}^{\dagger}\sigma_{F}),
\]
\[
\begin{split}
\dot\sigma_{F}&=-i\Omega\sigma_{F}{d}t-\dfrac{3}{4}\gamma\sigma_{F}dt\\&+\dfrac{\sqrt{\gamma}}{2}(\sigma_{R}a_{\rm{in}}+a_{\rm{in}}^{\dagger}\sigma_{R})+\sqrt{\gamma}(\sigma_{F}a_{\rm{in}}+a^{\dagger}_{\rm{in}}\sigma_{F}).
\end{split}
\]

From Sec.~\ref{sec::Strong drive} we know that in the steady state
$\braket{\sigma_{R,F,T}}=0$ when $c^{2}=s^{2}=1/2$. Averaging the equations above
then shows that correlators of the form $\braket{a_{\rm{in}}\sigma_{R,F,T}}$
vanish. This simplified argument is given for simultaneous correlators; the
general two-time case can be established using QSDE in Stratonovich or Ito
form for the operator $B\sigma_{R,F,T}$, where
$B(t)=\int_{t_{0}}^{t}a_{\rm{in}}(t^{\prime})dt^{\prime}$.

As a result, the normal reservoir correlator reduces to
\[
\braket{a_{\rm{out}}^{\dagger}(\omega_p)a_{\rm{out}}(\omega_{p^{\prime}})}=\gamma_{\rm{s}}\braket{\sigma_{R}^{\dagger}(\omega_{p})\sigma_{R}(\omega_{p^{\prime}})},
\]
and the anomalous one is
\[
\braket{a_{\rm{out}}(\omega_p)a_{\rm{out}}(\omega_{p^{\prime}})}=\gamma_{\rm{s}}\braket{\sigma_{R}(\omega_{p})\sigma_{R}(\omega_{p^{\prime}})}.
\]

Therefore, in Eq.~(\ref{squeezed part}) one can identify
$\braket{a_{p}^{\dagger}(t)a_{p^{\prime}}^{\dagger}(t^{\prime})}=F_{RR}(p,p^{\prime})$.

Consider the sum over momenta in (\ref{squeezed part}):
\[
\begin{split}
-e^{i\omega_{\rm{s}}(t+t^{\prime})}\sum_{p,p^{\prime}}\kappa_{p}\kappa_{p^{\prime}}e^{i\omega_{p}(t-t^{\prime})} F_{RR}(p,p^{\prime})=\\e^{i\omega_{\rm{s}}(t+t^{\prime})}\dfrac{\gamma_{\rm{pr}}}{2\pi}\int{d}\omega_{p}e^{i\omega_{p}(t-t^{\prime})} I_{Ri}\dfrac{1}{i\omega_{p}-\gamma_{0}}=\\e^{i\omega_{\rm{s}}(t+t^{\prime})}{\gamma_{\rm{pr}}}e^{-\gamma_{0}(t-t^{\prime})}I_{Ri}.
\end{split}
\]

The integrals of $F_{FT} (p,p^{\prime})$ in the calculation of the contribution are
suppressed as $\gamma_{\rm{s}}/\Omega_{\rm{s}}\ll1$. Physically, this reflects
that photons from the sidebands are off-resonant for the probe qubit in the
parameter regime considered here.

Here, the first Markov approximation was applied:
$\sum_{p}\kappa_{p_{\rm{s}}+p}\kappa_{p_{\rm{s}}-p}=\dfrac{\gamma_{\rm{pr}}}{2\pi}\int{d}\omega_{p}$,
where $\gamma_{\rm{pr}}$ is the radiative relaxation rate (linewidth) of the
probe qubit. $p_{s}$ is the momentum corresponding to $\omega_{\rm{s}}$.

The momentum integration produces a kernel with a characteristic time scale
$t_{\rm{ch}}=\gamma_{\rm{s}}^{-1}$. Taking into account that
$\sigma_{-}\rho(t^{\prime})\sigma_{-}\propto\braket{\sigma_{-}(t^{\prime})}$,
we note that the characteristic time of variation of
$\braket{\sigma_{-}(t^{\prime})}e^{i\omega_{\rm{pr}}t^{\prime}}$ is of order
$|\omega_{\rm{s}}-\omega_{\rm{pr}}|^{-1}\gg\gamma_{\rm{s}}^{-1}$. Hence, in
(\ref{squeezed part}) one can approximate
$\sigma_{-}\rho(t^{\prime})\sigma_{-}e^{i\omega_{\rm{pr}}t^{\prime}}\approx\sigma_{-}\rho(t)\sigma_{-}e^{i\omega_{\rm{pr}}t}$
(Markov approximation). Then (\ref{squeezed part}) takes the form
\[
\dot\rho\propto-{e}^{2i\omega_{\rm{s}}t}\gamma_{\rm{pr}}{M}\sigma_{-}\rho\sigma_{-},
\]
\begin{equation}\label{M expr}
M=2I_{Ri}\int\limits_{0}^{\infty}d\tau e^{-\gamma_{0}\tau}=\dfrac{2I_{Ri}}{\gamma_{0}}.
\end{equation}

For resonant drive on the source qubit, when $c^{2}=s^{2}=1/2$, $M=1$. Similarly,
from the term of (\ref{sec3::master eq general}) with $i=-$, $j=+$ it can be
shown that
\[
\dot{\rho}\propto-\dfrac{\gamma_{\rm{pr}}}{2}N\sigma_{+}\sigma_{-}\rho
\]
with $N=M=2I_{Ri}/\gamma_{0}$ calculated from the corresponding contribution in
(\ref{sec3::master eq general}).

The equality $N=M$ originates from the telegraph-like nature of the central component of the triplet. Two dressed-state-preserving processes form the central fluorescence component, $\ket{1}\rightarrow\ket{1}$ and $\ket{2}\rightarrow\ket{2}$. Their field amplitudes have equal magnitudes and opposite signs (\ref{sec2::Pauli new variables}). In the strong resonant-driving limit the two dressed states are equally populated, so the amplitudes cancel in the mean field, suppressing the coherent central line. The R component of the triplet can be treated as a symmetric random telegraph process $X(t)\rightarrow-X(t)$. However, $N$ adds up intensities, while $M$ adds up squared amplitudes (which relative phase is fixed at $\pi$) of the two dressed-state-preserving processes, both contributions add constructively in the second moments, yielding $N=M$.

The parameter \(N\) has a direct interpretation as the effective occupation of
the incident field sampled by the probe. The Markovian value is
\begin{equation}
N=2\,{\rm Re}\int_{0}^{\infty}d\tau\,
\left\langle
a_{\rm out}^{\dagger}(t+\tau)
a_{\rm out}(t)
\right\rangle.
\label{eq:N-occupation}
\end{equation}

For a probe with a
finite linewidth, the field correlations are sampled through its temporal
response function \(\chi_{\rm pr}(\tau)\), which decays on the time scale
\(\gamma_{\rm pr}^{-1}\). A corresponding finite-bandwidth occupation can therefore be defined as
\begin{equation}
N=
2\,{\rm Re}\int_{0}^{\infty}d\tau\,
\chi_{\rm pr}(\tau)
\left\langle
a_{\rm out}^{\dagger}(t+\tau)
a_{\rm out}(t)
\right\rangle .
\label{eq:N-filtered}
\end{equation}

Equivalently,
\begin{equation}
N
=
\int\frac{d\nu}{2\pi}\,
W_{\rm pr}(\nu)\,
S_{\rm out}(\omega_{\rm pr}+\nu),
\label{eq:N-spectral}
\end{equation}
where \(S_{\rm out}\) is the normally ordered spectrum of the incident
field and \(W_{\rm pr}\) is the spectral filter associated with the probe response. For a weak resonant probe,
\(\chi_{\rm pr}(\tau)=e^{-\gamma_{\rm pr}\tau/2}\), so that
\[
W_{\rm pr}(\nu)=
\frac{\gamma_{\rm pr}}
{\nu^{2}+(\gamma_{\rm pr}/2)^{2}},
\qquad
\int\frac{d\nu}{2\pi}W_{\rm pr}(\nu)=1 .
\]

Thus, \(N\) is the mean occupation of the incident radiation averaged
over the frequency window of width \(\sim\gamma_{\rm pr}\) selected by the
probe, rather than the total photon number in the complete source spectrum.

Thus, the master equation for the probe reads
\begin{equation}\label{sec3::master eq explicit}
\begin{split}
\dot\rho=&-\frac{i}{\hbar}\left[H_{\rm{sys}},\rho\right]+\dfrac{\gamma_{\rm{pr}}}{2}(N+1)(2\sigma_{-}\rho\sigma_{+}-\sigma_{+}\sigma_{-}\rho-\rho\sigma_{+}\sigma_{-})+\\& \dfrac{\gamma_{\rm{pr}}}{2}N(2\sigma_{+}\rho\sigma_{-}-\sigma_{-}\sigma_{+}\rho-\rho\sigma_{-}\sigma_{+})-\\&\ \frac{\gamma_{\rm{pr}}}{2}{M}e^{i\omega_{\rm{s}}(t+t^{\prime})}(\sigma_{+}\rho\sigma_{+}-\sigma_{+}\sigma_{+}\rho-\rho\sigma_{+}\sigma_{+})-\rm{h.c.}
\end{split}
\end{equation}

From (\ref{sec3::master eq explicit}) we obtain the equations of motion for the
probe averages:
\begin{widetext}
\begin{equation}\label{sec3::pauli eq1}
\braket{\dot\sigma_{-}}=\braket{\sigma_{-}}\left(-i\omega_{0}-\gamma_{\rm{pr}}\left(N+\dfrac{1}{2}\right)\right)-\dfrac{i\Omega_{\rm{pr}}}{2}e^{-i\omega_{\rm{pr}}t}\braket{\sigma_{-}}-\gamma_{\rm{pr}}{M}e^{-2i\omega_{\rm{s}}t}\braket{\sigma_{+}},
\end{equation}
\begin{equation}\label{sec3::pauli eq2}
    \braket{\dot\sigma_{z}}=-\gamma_{\rm{pr}}(\braket{\sigma_{z}}+1)-2N\gamma_{\rm{pr}}\braket{\sigma_{z}}+{i\Omega_{\rm{pr}}}(\braket{\sigma_{+}}e^{-i\omega_{\rm{pr}}t}-\braket{\sigma_{-}}e^{i\omega_{\rm{pr}}t}),
\end{equation}

\begin{figure*}[!pt]
    \centering
    \includegraphics[width=170mm, height=50mm]{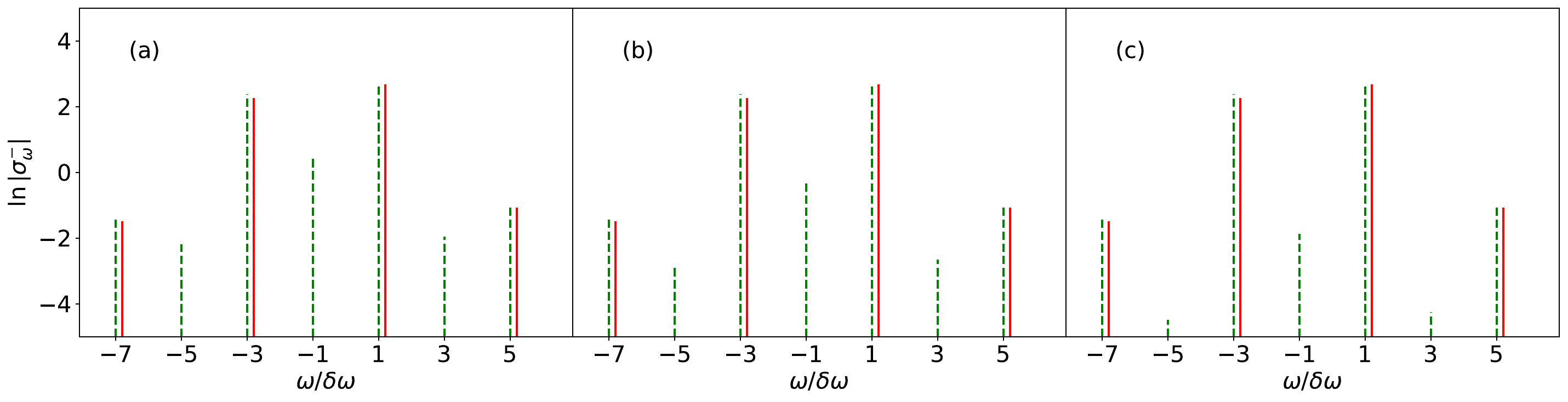} 
    \caption{Stationary QWM side-peak amplitudes for different ratios $\Omega_{\rm{s}}/\gamma_{\rm{s}}$:
    (a) $\Omega_{\rm{s}}/\gamma_{\rm{s}}=50$,
    (b) $\Omega_{\rm{s}}/\gamma_{\rm{s}}=100$,
    (c) $\Omega_{\rm{s}}/\gamma_{\rm{s}}=500$.
    Red solid line --- analytical solution, green dotted line --- numerical simulation. $\delta\omega=2\pi/100$, $\gamma_{\rm{pr}}/\gamma_{\rm{s}}=10$, $ \Omega_{\rm{pr}}/\gamma_{\rm{pr}}={0.2}$.}    \label{fig:peaks}
\end{figure*}

or, in the rotating frame characterized by the frequency
$\omega_{d}=(\omega_{\rm{s}}+\omega_{\rm{pr}})/2$,
$\delta\omega=\omega_{\rm{pr}}-\omega_{d}=\omega_{d}-\omega_{\rm{s}}$:
\begin{equation}\label{sec3::pauli eq1 rotating}
    \braket{\dot\sigma_{-}}=\gamma_{\rm{pr}}\left(N+\dfrac{1}{2}\right)\braket{\sigma_{-}}-\dfrac{i\Omega_{\rm{pr}}}{2}e^{-i\delta\omega t}\braket{\sigma_{-}}-\gamma_{\rm{pr}}{M}e^{2i\delta\omega t}\braket{\sigma_{+}},
\end{equation}
\begin{equation}\label{sec3::pauli eq2 rotating}
    \braket{\dot\sigma_{z}}=-\gamma_{\rm{pr}}(\braket{\sigma_{z}}+1)-2N\gamma_{\rm{pr}}\braket{\sigma_{z}}+{i\Omega_{\rm{pr}}}(\braket{\sigma_{+}}e^{-i\delta\omega t}-\braket{\sigma_{-}}e^{i\delta\omega t}).
\end{equation}
\end{widetext}

A coherent term from (\ref{sec2::I_R solution}), (\ref{sec2::RR anomal}) is
vanishing in the limit $\Omega_{\rm{s}}\gg\Delta$.

As shown in \cite{effects}, in the limit of weak pumping of the probe qubit
$\Omega_{\rm{pr}}\ll\gamma_{\rm{pr}}$ the solution can be written as
\begin{equation}\label{sec3::sigmam solution}
\begin{split}
\braket{\sigma_{-}}\approx&\dfrac{if\gamma_{\rm{pr}}}{\Omega_{\rm{pr}}}(fe^{i\delta\omega t}+fme^{-3i\delta\omega t}-\\&f^{3}m^{*}e^{5i\delta\omega t}-f^{3}m^{2}e^{-7i\delta\omega t}+...),
\end{split}
\end{equation}
where
\[
\begin{split}
&f=\dfrac{\Omega_{\rm{pr}}}{\gamma_{\rm{pr}}\sqrt{(2N+1)^2-4|M|^{2}}},\\
&m=\dfrac{2M}{2N+1}.
\end{split}
\]

When deriving the master equation (\ref{sec3::master eq general}), the Born
approximation was used, i.e.,
\[
\rho_{\rm{tot}}(t)\approx\rho(t){R_{0}}, 
\]
where $R_{0}$ is the density matrix of the reservoir at time zero, and
$\rho_{\rm{tot}}$ is the density matrix of the reservoir together with the probe
qubit system. This approximation assumes that the interaction with the probe
only weakly perturbs the reservoir.

\section{Discussion}\label{sec::Discussion}

\begin{figure}
\includegraphics[width=7cm, height=3cm]{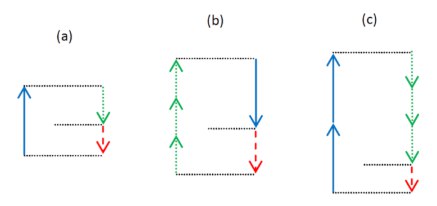} 
    \caption{\raggedright Schematic representation of multiphoton processes generating QWM side peaks at (a) $-3\delta\omega$, (b) $5\delta\omega$, and (c) $-7\delta\omega$. Blue arrows (solid lines) indicate absorption and emission of correlated photon pairs with total energy $2\omega_{\rm{s}}$. Green dotted arrows correspond to single photons of the coherent drive at frequency $\omega_{\rm{pr}}$. Red dashed arrows indicate photon emission responsible for the side peaks.}\label{fig conservation energy}
\end{figure}

Results of numerical simulations over a wide range of drive amplitudes for both
the probe and the source are summarized in Fig.~\ref{fig:color maps} of Appendix~\ref{app::finite-bandwidth}. In the
regime of strong source driving ($\Omega_{s}\gg\gamma_{s}$), the peaks at
frequencies $-5\delta\omega$,  $-1\delta\omega$, $+3\delta\omega$ are
suppressed, while peaks at other frequencies remain and become nearly
independent of the source drive. This behavior is consistent with the presence
of a squeezed component (or, equivalently, correlated photon pairs) in the
radiation of the strongly driven source. As discussed in
Sec.~\ref{sec::Strong drive}, the source emits photons in correlated pairs with
total energy $2\omega_{\rm{s}}$, and therefore QWM sidebands can only be
generated via processes involving absorption/emission of an even number of
photons originating from the source field. This selection rule is schematically
illustrated in Fig.~\ref{fig conservation energy}. Pair statistics of squeezed
light have been experimentally demonstrated in \cite{PhysRevA.81.013814}. The
weak-drive regime for the source qubit (left side of the plots) is discussed in
\cite{PhysRevA.111.043715}.

Numerical simulation is based on the full system of equations for the cascaded
setup in the spirit of \cite{Gardinder_cascaded}:
\begin{widetext}
\[
\dfrac{\partial\braket{{\sigma}_{-}^{\rm{pr}}}}{\partial{t}} = {\Omega_{\rm{pr}}}\braket{\sigma_{z}^{\rm{pr}}} e^{- i \delta\omega t} - \frac{\gamma_{\rm{pr}}}{2}\braket{\sigma_{-}^{\rm{pr}}}+ \mu\sqrt{\gamma_{\rm{pr}}\gamma_{\rm{s}}}\braket{\sigma_{-}^{\rm{s}}\sigma_{z}^{\rm{pr}}} ,
\]
\[
\dfrac{\partial\braket{{\sigma}_{z}^{\rm{pr}}}}{\partial t} = - \left(2 \Omega_{\rm{pr}}\braket{\sigma_{+}^{\rm{pr}}} e^{- i \delta\omega t}+ \text{h.c.} \right) - 2 \mu\sqrt{\gamma_{s}\gamma_{\rm{pr}}}( \braket{\sigma_{+}^{\rm{s}}\sigma_{-}^{\rm{pr}}} + \text{h.c.}) - \gamma_{\rm{pr}}(\braket{\sigma_{z}^{\rm{pr}}} + 1),
\]

\begin{figure*}[t]
    \centering
    \includegraphics[width=0.92\textwidth]{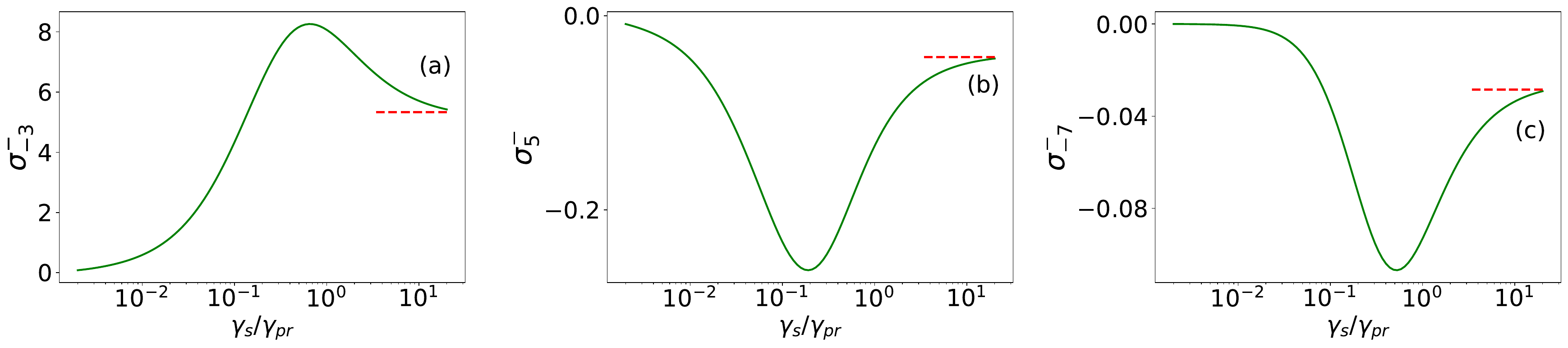}
    \caption{\raggedright
    Stationary Fourier amplitudes $\bar{\sigma}_{n}^{-}$ of representative QWM
    components as functions of the linewidth ratio
    $r=\gamma_{\rm{s}}/\gamma_{\rm{pr}}$ at fixed
    $\Omega_{\rm{s}}/\gamma_{\rm{s}}=100$.
    The green solid curves are obtained from the full cascaded two-qubit
    equations, while the red dashed lines show the analytical broadband values
    corresponding to $N=M=1$. The panels correspond to
    (a) $n=-3$, (b) $n=5$, (c) n=$-7$.
    The remaining parameters are
    $\Omega_{\rm{pr}}/\gamma_{\rm{pr}}=0.1$ and
    $\delta\omega/\gamma_{\rm{pr}}=2\pi/100$.}
    \label{fig:gamma-dependence}
\end{figure*}
together with the equations on the simultaneous correlators like $\braket{\sigma_{+}^{\rm{s}}\sigma_{z}^{\rm{pr}}}$, $\braket{\sigma_{-}^{\rm{s}}\sigma_{z}^{\rm{pr}}}$ etc. Details are discussed in Appendix \ref{app::Gardiner}. The amplitude of the elastically scattered wave is $-i\braket{\sigma_{-}}/\mu$,
where $\mu$ is the qubit dipole moment \cite{Astafiev2010resonance}.
\end{widetext}

The analytical solution also shows explicitly which properties of the incident
field are encoded in the QWM spectrum. Let $\bar{\sigma}_{n}^{-}$ denote the
complex Fourier amplitude at
$\omega_n=\omega_d+n\delta\omega$. In the weak-probe broadband regime,
Eq.~(\ref{sec3::sigmam solution}) gives
\begin{equation}
m=\frac{\bar{\sigma}_{-3}^{-}}{\bar{\sigma}_{1}^{-}},
\qquad
f^{2}=
\left|
\frac{\bar{\sigma}_{5}^{-}}{\bar{\sigma}_{-3}^{-}}
\right|.
\label{eq:fm-from-peaks}
\end{equation}
Using $m=2M/(2N+1)$ and $f=\Omega_{\rm{pr}}/[\gamma_{\rm{pr}}\sqrt{(2N+1)^2-4|M|^2}]$,
one obtains
\begin{equation}
N=
\frac{1}{2}
\left[
\frac{\Omega_{\rm{pr}}}
{\gamma_{\rm{pr}}f\sqrt{1-|m|^{2}}}
-1
\right],\ \ 
M=
\frac{m}{2}
\frac{\Omega_{\rm{pr}}}
{\gamma_{\rm{pr}}f\sqrt{1-|m|^{2}}}.
\label{eq:NM-from-peaks}
\end{equation}

Thus, within the effective squeezed-field description, the relative amplitudes
of several QWM peaks determine the normal and anomalous correlation parameters
of the field incident on the probe.

The solution of Eqs.~(\ref{sec3::pauli eq1 rotating})--(\ref{sec3::pauli eq2 rotating})
(in the stationary approximation, taking into account that $\delta\omega{t}$ is
a slowly varying phase on the time scale $\gamma_{\rm{pr}}^{-1}$) at a fixed
large drive on the source is shown in Fig.~\ref{fig:peaks} together with the
numerical simulations. The resulting spectrum coincides with the QWM spectrum
for squeezed light at the output of a degenerate parametric amplifier mixed
with a coherent signal for $N=1$ and $M=1$, as discussed in \cite{effects}.

To quantify the range of validity of the broadband description, we now keep the
source in the same strongly saturated regime,
$\Omega_{\rm{s}}/\gamma_{\rm{s}}=100$, and vary the linewidth ratio
\[
r=\frac{\gamma_{\rm{s}}}{\gamma_{\rm{pr}}}.
\]
In this way, the normalized stationary state of the source remains practically
unchanged, whereas the correlation time of its fluorescence is varied relative
to the response time of the probe. Figure~\ref{fig:gamma-dependence} compares
the Fourier amplitudes of QWM components $-3\delta\omega,\ 5\delta\omega,\ -7\delta\omega$ obtained from the
full cascaded equations with their analytical values in the broadband limit.

For $r\gg1$, numerical amplitudes approach the analytical broadband values. However, the approach to this limit is not monotonic. Upon decreasing $r$, the amplitudes of the allowed QWM sidebands first increase above their broadband
values and only then vanish in the narrow-source limit. The maxima of the
$n=-3$ and $n=-7$ components occur near $r\simeq0.6$, whereas the $n=5$
component reaches its largest magnitude near $r\simeq0.2$. The maximum
magnitude is approximately $1.5$ times the broadband value for the $n=-3$
component and is enhanced by factors of about $6$ and $4$ for the $n=5$ and
$n=-7$ components, respectively.

The nonmonotonic behavior can be understood from a simple finite-bandwidth
estimate. For resonant strong driving, the normal and anomalous correlators of
the incoherent central fluorescence component have the same exponential
envelope $\propto{e}^{-\gamma_{0}|\tau|}$. 

The probe cannot respond equally efficiently to the entire correlation function
when its own coherence time becomes comparable to $\gamma_{0}^{-1}$. As qualitively shown in Appendix \ref{appsec::overlap_estimate}, $N$ and $M$ can be estimated 
\[
N_{\rm eff} = M_{\rm eff}\simeq \frac{r}{1+r}.
\]

This replacement is not intended as a Markovian master equation for
$r\sim1$; it only estimates the reduction of the normal and anomalous
correlation weights sampled by the probe.

Such estimation predicts quantitively the maxima at intermediate values of $r$ which can be seen in Appendix \ref{appsec::overlap_estimate}:
\[
r_{5}^{\rm max}\simeq0.18,
\qquad
r_{-3}^{\rm max}=r_{-7}^{\rm max}\simeq0.55
\]
in reasonable agreement with the numerical maxima near $0.2$ and $0.6$, respectively.

Decreasing the source bandwidth therefore has two competing consequences.
It reduces the pair-correlation parameter $m$ (\ref{sec3::sigmam solution}), which tends to suppress the QWM
sidebands, but it simultaneously reduces the normal source noise and the
source-induced damping and partial saturation of the probe. The latter effect
increases the nonlinear probe susceptibility through the factor $f$. Their
competition produces the maxima at intermediate linewidth ratios.

Although the absolute QWM amplitudes differ substantially from the broadband
prediction at $r\sim1$, the selection rule remains visible in the complete
numerical maps shown in Fig.~\ref{fig:color maps} in Appendix $\ref{app::finite-bandwidth}$. This robustness follows
from a symmetry of the central dressed-state scattering channel and does not require a white-noise approximation, only a strong drive and near-resonance, $\Omega_{\rm{s}}\gg\gamma_{\rm{s}},\Delta$. In this limit the R component is a symmetric telegraph process $X(t)\rightarrow{-X}(t)$ (see discussion above Eq. \ref{sec3::master eq explicit}) and its odd multitime correlators vanish,
\[
\left\langle
X(t_{1})X(t_{2})\cdots X(t_{2k+1})
\right\rangle=0.
\]

Consequently, there are no multiphoton processes on the probe that involve an odd number of photons from the source on the frequency $\omega_{\rm{s}}$. Corresponding QWM contributions cancel, independently of whether the correlation time is short or
comparable to the probe response time.

In order to apply the Born approximation and to use the Lindblad equation for
the source qubit, it was assumed that $\gamma_{\rm{pr}}\ll\gamma_{\rm{s}}$. This
regime corresponds to filtering the broad source spectrum by a narrow filter
implemented by the probe qubit in a frequency window near $\omega_{\rm{s}}$
(the center of the fluorescence triplet). The filtered intensity correlation function
of resonant fluorescence in the narrow-filter limit $\lambda\ll\gamma_{\rm{s}}$,
where $\lambda=\gamma_{\rm{pr}}$ is the filter width, has the form \cite{Nienhuis_1}:
\[
\begin{split}
g_{2}(\omega_{1},\omega_{2}, t)=1+&\Big(\dfrac{4\lambda^2}{4\lambda^{2}+(\omega_{1}-\omega_{2})^2}+\\ &\dfrac{4\lambda^2}{4\lambda^{2}+(\omega_{1}+\omega_{2})^2}\Big)e^{-\lambda{t}}.
\end{split}
\]

Thus, for frequencies symmetrically located relative to $\omega_{\rm{s}}$,
photon bunching is observed. This bunching provides evidence that photons are
emitted in pairs. At maximum, $g_2(\omega_{1},\omega_{2}, 0)$ is equal to $2$.
In this narrowband regime the probe qubit effectively detects squeezed
radiation (correlated photon pairs) with an anomalous correlator $M=1$, as well
as thermal broadband radiation (photons at the same frequency) with $N=1$, where
$N$ and $M$ are determined by
\[
\begin{split}
&\braket{a_{\text{in}}(t_{1})a_{\text{in}}(t_{2})}=M\delta(t_{1}-t_{2})e^{-2\omega_{\rm{s}}(t_{1}+t_{2})},\\
&\braket{a_{\text{in}}^{\dagger}(t_{1})a_{\text{in}}(t_{2})}=N\delta(t_{1}-t_{2}).
\end{split}
\]

In the opposite limit $\gamma_{\rm{pr}}\gg\gamma_{\rm{s}}$, all peaks except the
peak at $\omega_{\rm{pr}}$ are suppressed, as can be seen in the bottom row of
Fig.~\ref{fig:color maps} in Appendix \ref{app::finite-bandwidth}. As shown in \cite{Nienhuis_2}, in the wide-filter
limit ($\lambda\gg\gamma_{\rm{s}}$) photon correlations within the $R$ peak
vanish, $g_{2}(\omega_{1}, \omega_{2}, t)\approx1$ (if $\omega_{1}$, $\omega_{2}$
belong to the $R$ peak). The dependence of the intensity correlation function on
the filter width at different pumping conditions has been experimentally
demonstrated in \cite{PhysRevLett.125.043603}.

The cascaded architecture considered here has already been
realized experimentally using tunable transmon qubits with transition
frequencies near \(5.1~\mathrm{GHz}\)
\cite{PhysRevA.111.043715}. In that experiment, the radiative linewidths were
\(\gamma_{\rm{s}}/2\pi\simeq1.7~\mathrm{MHz}\) and
\(\gamma_{\rm{pr}}/2\pi\simeq1.8~\mathrm{MHz}\), the corresponding
pure-dephasing rates were approximately \(0.15~\mathrm{MHz}\) and
\(0.19~\mathrm{MHz}\), and the field-amplitude transmission between the source
and the probe was about \(0.79\). These parameters realize the intermediate
regime \(\gamma_{\rm{s}}/\gamma_{\rm{pr}}\sim1\), in which our calculations
predict that the selection rule remains visible although the broadband
approximation is no longer quantitatively accurate. The regime
\(\gamma_{\rm{s}}\gg\gamma_{\rm{pr}}\) can be approached by engineering
asymmetric qubit--waveguide couplings. For example, a source linewidth of a few
megahertz, a probe linewidth of a few tenths of a megahertz, and a source Rabi
frequency of several tens of megahertz would satisfy both
\(\Omega_{\rm{s}}\gg\gamma_{\rm{s}}\) and
\(\gamma_{\rm{s}}\gg\gamma_{\rm{pr}}\), while remaining compatible with a
typical transmon anharmonicity of a few hundred megahertz. Pure dephasing should
remain smaller than the probe radiative linewidth, whereas propagation losses
primarily reduce the source-induced QWM amplitudes and can be included through
the source-to-probe transmission coefficient.

\section{Conclusion}

We have developed a microscopic description of QWM in a cascaded
two-qubit waveguide-QED system, in which the probe qubit is driven
simultaneously by an external coherent tone and by the resonance fluorescence
of a strongly driven source qubit. Starting from the two-time correlation
functions of the source emission, we separated the coherent and fluctuating
parts of the central component of the fluorescence triplet and identified its
normal and anomalous correlators. For strong near-resonant driving, the coherent
Rayleigh contribution is suppressed, whereas the remaining central component
retains phase-sensitive pair correlations satisfying the frequency constraint
$\omega+\omega'=2\omega_{\rm{s}}$. In the broadband-source, or narrow-probe,
limit $\gamma_{\rm{s}}\gg\gamma_{\rm{pr}}$, these correlations lead to an
effective master equation identical in form to that of a qubit driven by a
coherent tone and broadband squeezed radiation.

The comparison with Ref.~\cite{effects} also identifies the physical content
specific to the present cascaded realization. In Ref.~\cite{effects}, the
normal and anomalous reservoir parameters were prescribed as properties of an
external nonclassical input. Here, by contrast, the complete
resonance-fluorescence field emitted by the source qubit remains antibunched
and strongly non-Gaussian. Only after spectral selection by the probe does its
central component admit an effective Gaussian description in terms of the
normal and anomalous correlators \(N\) and \(M\). Their effective values are
determined not only by the source dynamics, but also by the spectral window
sampled by the probe. In this sense, \(N\) represents the occupation of the
incident field within the probe bandwidth, while \(M\) characterizes
phase-sensitive pair correlations in the same frequency window. The cascaded geometry therefore combines the generation, spectral filtering,
and nonlinear readout of nonclassical radiation within a two-qubit system. In
the broadband regime, the source qubit plays the role of the effective
squeezed-field source and removes the need for a separate DPA, thereby
simplifying a possible experimental implementation of QWM with nonclassical
radiation.

We have also examined the finite-bandwidth regime by varying the linewidth ratio
$\gamma_{\rm{s}}/\gamma_{\rm{pr}}$. The QWM selection rule remains visible even
for $\gamma_{\rm{s}}\sim\gamma_{\rm{pr}}$, since it originates from the sign
symmetry of the central dressed-state fluorescence component rather than from
the Markov approximation itself. Moreover, the allowed sidebands exhibit a
nonmonotonic dependence on the linewidth ratio and notably exceed their broadband
values at intermediate ratios. This enhancement results from the competition
between the weakening of the pair-correlated source field and the simultaneous
reduction of source-induced probe broadening, which increases the nonlinear
susceptibility of the probe. In the opposite limit
$\gamma_{\rm{s}}\ll\gamma_{\rm{pr}}$, the spectral overlap becomes small, and
the nontrivial QWM sidebands are suppressed.

Finally, the QWM spectrum contains quantitative information about the
correlations of the field incident on the probe. In the weak-probe broadband
regime, ratios of selected sideband amplitudes determine the normal and anomalous correlators $N$ and $M$.
For the cascaded system considered here, the selection rule and the amplitudes
of the surviving sidebands therefore provide a spectral characterization of
the pair-correlated central component of the source fluorescence. QWM thus
offers a correlator-sensitive readout of nonclassical radiation generated by a
coherently driven artificial atom.

\section{Acknowledgements}
We thank E. S. Andrianov for very useful discussions. 
The study is supported by the Ministry of Science and Higher Education of the Russian Federation (agreement No.
075-15-2024-538).

\appendix

\section{Full system of equations of a cascaded system}\label{app::Gardiner}

In this Appendix we summarize the full set of equations used for numerical
simulations of the cascaded source--probe system. Following
\cite{Gardinder_cascaded, PhysRevLett.70.2269}, we write the master equation in
the form
\begin{equation}\label{app::master_eq Gardiner}
\begin{split}
\dfrac{d\rho}{dt}=&\dfrac{i}{\hbar}[\rho, H_{\rm{sys}}]-\\&\sqrt{\gamma_{\rm{pr}}\gamma_{\rm{s}}}\mu\left([\sigma^{\rm{pr}_{+}},\sigma^{\rm{s}}_{-}\rho]+[\rho\sigma^{\rm{s}}_{+},\sigma^{\rm{pr}}_{-}]\right)+\\&\hat{L}_{1}\rho+\hat{L}_{2}\rho,
\end{split}
\end{equation}
where $\mu$ is the fraction of radiation passing between the source qubit and
the probe qubit,
\[
\begin{split}
H_{\rm{sys}}=&\dfrac{1}{2}\hbar\omega_{1}\sigma_{z}^{\rm{s}}+\dfrac{1}{2}\hbar\omega_{2}\sigma_{z}^{\rm{pr}}+\\&
\left(\Omega_{\rm{s}}e^{-i\omega_{\rm{s}}t}\sigma_{+}^{\rm{pr}}+\text{h.c.} \right)+\\&\left(\Omega_{\rm{pr}}e^{-i\omega_{\rm{pr}}t}\sigma_{+}^{\rm{pr}}+\text{h.c.} \right),
\end{split}
\]
where $\Omega_{\rm{s}}$, $\omega_{\rm{s}}$, $\omega_{1}$  and $\Omega_{\rm{pr}}$,
$\omega_{\rm{pr}}$, $\omega_{2}$ --- the Rabi frequency, drive frequency, and transition
frequency of the source qubit and probe qubit, respectively. For
simplicity, let us assume that $\omega_{1}=\omega_{2}=\omega_{0}$.

Expressions $\hat{L}_{1}\rho$ and $\hat{L}_{2}\rho$ are defined by
\[
\begin{split}
&\hat{L}_{1}\rho=\dfrac{1}{2}\gamma_{\rm{s}}\left(2\sigma_{-}^{\rm{s}}\rho\sigma_{+}^{\rm{s}}-\sigma_{+}^{\rm{s}}\sigma_{-}^{\rm{s}}\rho-\rho\sigma_{+}^{\rm{s}}\sigma_{-}^{\rm{s}}\right),\\&\hat{L}_{2}\rho=
\dfrac{1}{2}\gamma_{\rm{pr}}\left(2\sigma_{-}^{\rm{pr}}\rho\sigma_{+}^{\rm{pr}}-\sigma_{+}^{\rm{pr}}\sigma_{-}^{\rm{pr}}\rho-\rho\sigma_{+}^{\rm{pr}}\sigma_{-}^{\rm{pr}}\right).
\end{split}
\]

From the density matrix equation \ref{app::master_eq Gardiner}, one can obtain
equations for the averages. Equations for the source qubit and probe qubit
operators, written in the rotating frame characterized by the transition
frequency of the probe qubit, where time is dimensionless $\tau=\gamma_{pr}t$:
\begin{widetext}
\[
\dfrac{\partial\braket{{\sigma}_{-}^{\rm{pr}}}}{\partial{\tau}} = \frac{\Omega_{\rm{pr}}}{\gamma_{\rm{pr}}} \braket{\sigma_{z}^{\rm{pr}}} e^{- i \delta\omega t} + \alpha \braket{\sigma_{-}^{\rm{s}}\sigma_{z}^{\rm{pr}}} - \frac{\braket{\sigma_{-}^{\rm{pr}}}}{2},
\]
\[
\dfrac{\partial\braket{{\sigma}_{z}^{\rm{pr}}}}{\partial\tau} = - \left(\frac{2 \Omega_{\rm{pr}}}{\gamma_{\rm{pr}}} \braket{\sigma_{+}^{\rm{pr}}} e^{- i \delta\omega t}+ \text{h.c.} \right) - 2 \alpha( \braket{\sigma_{+}^{\rm{s}}\sigma_{-}^{\rm{pr}}} + \text{h.c.}) - \braket{\sigma_{z}^{\rm{pr}}} - 1,
\]
\[
\dfrac{\partial\braket{{\sigma_{-}^{\rm{s}}\sigma_{+}^{\rm{pr}}}}}{\partial\tau} = \frac{\Omega_{\rm{s}}}{\gamma_{\rm{pr}}}\braket{\sigma_{z}^{\rm{s}}\sigma_{+}^{\rm{pr}}} e^{i \delta\omega t} + \dfrac{\overline{\Omega_{\rm{pr}}}}{\gamma_{\rm{pr}}}\braket{\sigma_{-}^{\rm{s}}\sigma_{z}^{\rm{pr}}} e^{i \delta\omega t} + \alpha\left(\frac{ \braket{\sigma_{z}^{\rm{pr}}}}{2} + \frac{\braket{\sigma_{z}^{\rm{s}}\sigma_{z}^{\rm{pr}}}}{2}\right)- \braket{\sigma_{-}^{\rm{s}}\sigma_{+}^{\rm{pr}}} \left( \frac{\gamma_{\rm{s}}}{2 \gamma_{\rm{pr}}} + \frac{1}{2}\right), 
\]
\[
\dfrac{\partial\braket{{\sigma_{+}^{\rm{s}}\sigma_{+}^{\rm{pr}}}}}{\partial\tau} = -\braket{\sigma_{+}^{\rm{s}}\sigma_{+}^{\rm{pr}}} \left( \frac{\gamma_{\rm{s}}}{2 \gamma_{\rm{pr}}} + \frac{1}{2}\right) + \frac{\overline{\Omega_{\rm{pr}}}}{\gamma_{\rm{pr}}}\braket{\sigma_{+}^{\rm{s}}\sigma_{z}^{\rm{pr}}} e^{i \delta\omega t}  +\frac{\overline{\Omega_{\rm{s}}}}{\gamma_{\rm{pr}}} \braket{\sigma_{z}^{\rm{s}}\sigma_{+}^{\rm{pr}}} e^{- i \delta\omega t}, 
\]
\[
\begin{split}
\dfrac{\partial\braket{{\sigma_{+}^{\rm{s}}\sigma_{z}^{\rm{pr}}}}}{\partial\tau} = -& \left(\frac{2 \Omega_{\rm{pr}}}{\gamma_{\rm{pr}}}\braket{\sigma_{+}^{\rm{s}}\sigma_{+}^{\rm{pr}}} e^{- i \delta\omega t}+\frac{2\overline{\Omega_{\rm{pr}}}}{\gamma_{\rm{pr}}}\braket{\sigma_{+}^{\rm{s}}\sigma_{-}^{\rm{pr}}} e^{i \delta\omega t}\right)+ \frac{\overline{\Omega_{\rm{s}}}}{\gamma_{\rm{pr}}}\braket{\sigma_{z}^{\rm{s}}\sigma_{z}^{\rm{pr}}} e^{- i \delta\omega t}  - \\ &\alpha \braket{\sigma_{+}^{\rm{pr}}} - \alpha \braket{\sigma_{z}^{\rm{s}}\sigma_{+}^{\rm{pr}}}  - \braket{\sigma_{+}^{\rm{s}}\sigma_{z}^{\rm{pr}}} \left( \frac{\gamma_{\rm{s}}}{2 \gamma_{\rm{pr}}} + 1\right) - \braket{\sigma_{+}^{\rm{s}}},
\end{split}
\]
\[
\begin{split}
\dfrac{\partial\braket{{\sigma_{z}^{\rm{s}}\sigma_{z}^{\rm{pr}}}}}{\partial\tau} = &- \left(\frac{2 \Omega_{\rm{pr}}}{\gamma_{\rm{pr}}}\braket{\sigma_{z}^{\rm{s}}\sigma_{+}^{\rm{pr}}} e^{- i \delta\omega t}+\text{h.c.}\right)-\left(\frac{2 \Omega_{\rm{s}}}{\gamma_{\rm{pr}}} \braket{\sigma_{+}^{\rm{s}}\sigma_{z}^{\rm{pr}}} e^{i \delta\omega t}+\text{h.c.}\right)  +\\& 2 \alpha (\braket{\sigma_{+}^{\rm{s}}\sigma_{-}^{\rm{pr}}} + \text{h.c.})  -   \braket{\sigma_{z}^{\rm{pr}}}\frac{ \gamma_{\rm{s}}}{\gamma_{\rm{pr}}} -\braket{\sigma_{z}^{\rm{s}}\sigma_{z}^{\rm{pr}}} \left( \frac{\gamma_{\rm{s}}}{\gamma_{\rm{pr}}} + 1\right) - \braket{\sigma_{z}^{\rm{s}}},
\end{split}
\]
\[
\begin{split}
\dfrac{\partial\braket{{\sigma_{z}^{\rm{s}}\sigma_{-}^{\rm{pr}}}}}{\partial\tau} = \frac{\Omega_{\rm{pr}}}{\gamma_{\rm{pr}}} \braket{\sigma_{z}^{\rm{s}}\sigma_{z}^{\rm{pr}}} e^{- i \delta\omega t} -& \left(\frac{2 \Omega_{\rm{s}}}{\gamma_{\rm{pr}}} \braket{\sigma_{+}^{\rm{s}}\sigma_{-}^{\rm{pr}}} e^{i \delta\omega t}+ \frac{2 \overline{\Omega_{\rm{s}}}}{\gamma_{\rm{pr}}}\braket{\sigma_{-}^{\rm{s}}\sigma_{-}^{\rm{pr}}} e^{- i \delta\omega t} \right) - \\&\alpha \braket{\sigma_{-}^{\rm{s}}\sigma_{z}^{\rm{pr}}} - \braket{\sigma_{-}^{\rm{pr}}}\frac{ \gamma_{\rm{s}}}{\gamma_{\rm{pr}}}  -\braket{\sigma_{z}^{\rm{s}}\sigma_{-}^{\rm{pr}}} \left( \frac{\gamma_{\rm{s}}}{\gamma_{\rm{pr}}} + \frac{1}{2}\right),
\end{split}
\]
\end{widetext}
where $\alpha=\mu\sqrt{\dfrac{\gamma_{\rm{s}}}{\gamma_{\rm{pr}}}}$.

The system should be supplemented with equations for the source qubit:
\[
\frac{\partial\braket{{\sigma}_{-}^{\rm{s}}}}{\partial{t}} = {\Omega_{\rm{s}} e^{i \delta\omega t}}\braket{\sigma_{z}^{\rm{s}}} - \frac{\gamma_{s}}{2}\braket{\sigma_{-}^{\rm{s}}},
\]
\[
\frac{\braket{{\sigma}_{z}^{\rm{s}}}}{\partial t} = - 2\left( \Omega_{s} \braket{\sigma_{+}^{\rm{s}}} e^{i \delta\omega t} - \text{h.c.}\right) - \gamma_{s}(\braket{\sigma_{z}^{\rm{s}}}+1),
\]
which have the solution:
\[
\braket{\sigma_{z}^{\rm{s}}}=-\dfrac{1}{1+\dfrac{8|\Omega_{\rm{s}}|^2}{\gamma_{\rm{s}}^{2}}},
\]
\[
\braket{\sigma_{-}^{\rm{s}}}=-\dfrac{2\Omega_{\rm{s}}e^{i\delta\omega{t}}}{\gamma_{\rm{s}}\left(1+\dfrac{8|\Omega_{\rm{s}}|^2}{\gamma_{\rm{s}}^{2}}\right)}.
\]

\begin{figure*}[ht]
    \centering
    \includegraphics[width=\textwidth]{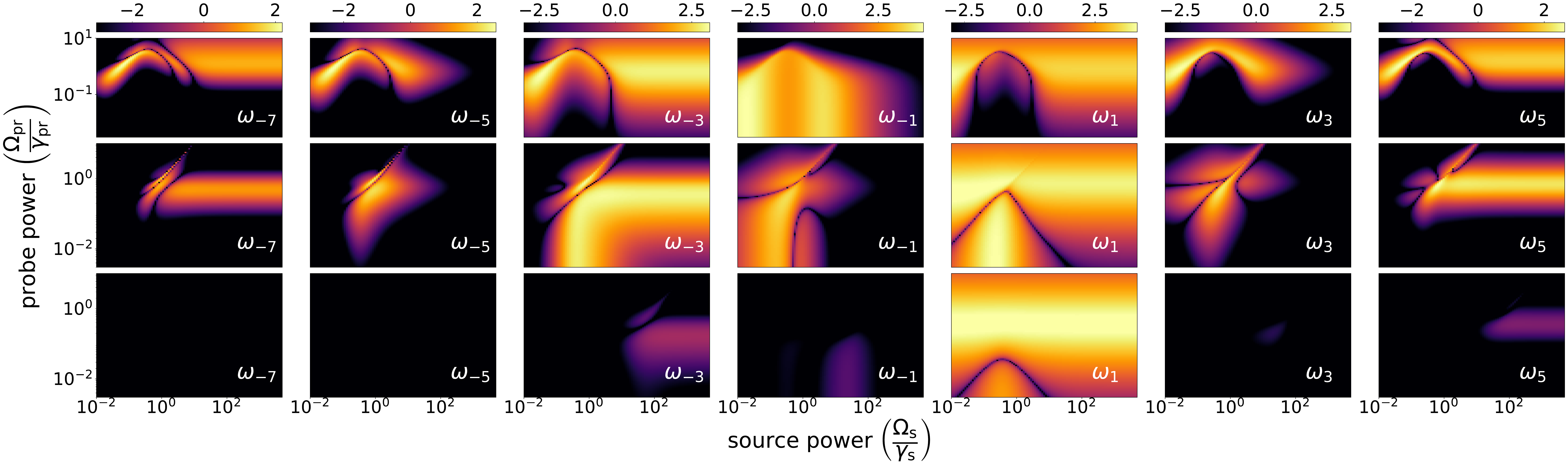}
    \caption{\raggedright
    Full numerical maps of the seven QWM side-peak intensities in the cascaded
    source--probe system. From left to right, the columns correspond to
    $n=-7,-5,-3,-1,1,3,$ and $5$, where
    $\omega_n=\omega_d+n\delta\omega$. Each row corresponds to a different
    linewidth ratio: upper row,
    $\gamma_{\rm{s}}/\gamma_{\rm{pr}}=100$; middle row,
    $\gamma_{\rm{s}}/\gamma_{\rm{pr}}=1$; and bottom row,
    $\gamma_{\rm{s}}/\gamma_{\rm{pr}}=0.01$.}
    \label{fig:color maps}
\end{figure*}

If $\delta\omega\ll\rm{min}(\gamma_{\rm{s}}, \gamma_{\rm{pr}})$, one can use the
stationary approximation, considering $\delta\omega{t}$ to be a slowly varying
phase, and solve the system as a system of linear algebraic equations.

\section{Full QWM maps and finite-bandwidth corrections}
\label{app::finite-bandwidth}

For completeness, Fig.~\ref{fig:color maps} presents the full set of
numerically calculated QWM maps that underlies the representative linewidth
dependences discussed in Sec.~\ref{sec::Discussion}. The three rows correspond
to the broadband-source, intermediate, and narrow-source regimes. The
suppressed branch $n=4l-1$, represented by the components
$n=-5,-1,3$, remains weak in all three cases. At
$\gamma_{\rm{s}}/\gamma_{\rm{pr}}=0.01$, the nontrivial components of the
allowed branch are also strongly reduced, but this suppression is not
accompanied by a transfer of spectral weight to the complementary branch.

\subsection{Lorentzian-overlap estimate}\label{appsec::overlap_estimate}

We now derive the finite-bandwidth estimate used in
Sec.~\ref{sec::Discussion}. For resonant strong driving,
$c^{2}=s^{2}=1/2$, the coherent part of the central fluorescence component
vanishes. In a frame rotating at $\omega_{\rm{s}}$, the incoherent normal and
anomalous correlators have identical envelopes
\begin{equation}
C_N(\tau)=C_M(\tau)
=I_{Ri}e^{-\gamma_{0}\tau},
\qquad \tau>0,
\label{app:eq:central-correlators}
\end{equation}
where
\[
\gamma_{0}=\frac{\gamma_{\rm{s}}}{2},
\qquad
I_{Ri}=\frac{\gamma_{\rm{s}}}{4}.
\]

The equality of the two envelopes is the origin of the relation $N=M$ in the broadband reduction.

In the Markov limit, the normal and anomalous coefficients entering the probe master equation are 
\[
N=M=2I_{Ri}\int_{0}^{\infty}d\tau\,e^{-\gamma_{0}\tau}
=\frac{2I_{Ri}}{\gamma_{0}}=1.
\]
For finite $\gamma_{0}$, the probe operator cannot be regarded as constant over
the entire memory time. To lowest order, we approximate the relevant
transverse probe mode by
\[
\begin{split}
&e^{i\omega_{\rm{pr}}(t-\tau)}\sigma_{-}\rho(t-\tau)\sigma_-\simeq
e^{-(\gamma_{\rm{pr}}/2)\tau}e^{i\omega_{\rm{pr}}t}\sigma_{-}\rho(t)\sigma_-,\\
&e^{i\omega_{\rm{pr}}\tau}\sigma_+\sigma_-\rho(t-\tau)\simeq
e^{-(\gamma_{\rm{pr}}/2)\tau}\sigma_+\sigma_-\rho(t),
\end{split}
\]
compare with the paragraph before (\ref{M expr}).

The corresponding memory integral becomes
\begin{equation}
2I_{Ri}\int_{0}^{\infty}d\tau\,
e^{-(\gamma_{0}+\gamma_{\rm{pr}}/2)\tau}
=
\frac{2I_{Ri}}{\gamma_{0}+\gamma_{\rm{pr}}/2}.
\label{app:eq:memory-integral}
\end{equation}

Using $2I_{Ri}=\gamma_{0}$ gives
\begin{equation}\label{app:eq:q}
N_{\rm eff}\simeq M_{\rm eff}\simeq q,
\qquad
q=\frac{\gamma_{0}}
{\gamma_{0}+\gamma_{\rm{pr}}/2}
=\frac{r}{1+r},
\end{equation}
where $r=\gamma_{\rm{s}}/\gamma_{\rm{pr}}$.

Equation~(\ref{app:eq:q}) can equivalently be viewed as the overlap between the
Lorentzian central fluorescence line and the finite response bandwidth of the
probe. It preserves $N_{\rm eff}=M_{\rm eff}$ because the normal and anomalous
source correlators have the same temporal envelope; finite bandwidth reduces
their common weight.

Eq.~(\ref{app:eq:q}) should be
regarded as a one-parameter estimate rather than as a complete reduced master
equation outside the broadband limit. The master equation is not strictly valid beyond the limit $\gamma_{\rm{s}}\gg\gamma_{\rm{pr}}$, because the Markovian and Born approximations are violated. 

\subsection{Positions of the maxima}

Let
\[
\epsilon=\frac{\Omega_{\rm{pr}}}{\gamma_{\rm{pr}}}.
\]
With $N=M=q$, the parameters entering
Eq.~(\ref{sec3::sigmam solution}) become
\begin{equation}
f=\frac{\epsilon}{\sqrt{1+4q}},
\qquad
m=\frac{2q}{1+2q}.
\label{app:eq:fmq}
\end{equation}
Writing the probe coherence as
\[
\langle\sigma_-(t)\rangle
=\sum_n\bar{\sigma}_{n}^{-}e^{in\delta\omega t},
\]
Eq.~(\ref{sec3::sigmam solution}) gives, up to the common phase convention,
\begin{equation}
\begin{split}
\bar{\sigma}_{1}^{-}
&=
i\epsilon\frac{1}{1+4q},\\
\bar{\sigma}_{-3}^{-}
&=
i\epsilon
\frac{2q}{(1+4q)(1+2q)},\\
\bar{\sigma}_{5}^{-}
&=
-i\epsilon^{3}
\frac{2q}{(1+4q)^{2}(1+2q)},\\
\bar{\sigma}_{-7}^{-}
&=
-i\epsilon^{3}
\frac{4q^{2}}
{(1+4q)^{2}(1+2q)^{2}}.
\end{split}
\label{app:eq:four-amplitudes}
\end{equation}

The directly driven component $\bar{\sigma}_{1}^{-}$ decreases monotonically
with $q$. Thus, as the source becomes narrower and $q\rightarrow0$, this
component approaches the bare coherent-probe response.

Using (\ref{app:eq:four-amplitudes}), we find the points of maximum over $q=r/(1+r)$ of the $-3,\ 5$ and  $-7$ components:
\begin{equation}
q_{-3}^{\rm max}=q_{-7}^{\rm max}=\frac{1}{2\sqrt{2}},\ \ q_{5}^{\rm max}
=\frac{\sqrt{5}-1}{8},
\label{app:eq:qmax3}
\end{equation}

Using $r=q/(1-q)$, one obtains
\begin{equation}
r_{-3}^{\rm max}=r_{-7}^{\rm max}
\simeq0.55,\ \ r_{5}^{\rm max}
\simeq0.18.
\label{app:eq:rmax37}
\end{equation}

These estimates reproduce both the nonmonotonic character of the numerical
curves and the fact that the $n=5$ component reaches its maximum at a
substantially smaller linewidth ratio than the $n=-3$ and $n=-7$ components.

Despite the fact that intermediate values of $r$ are outside the range of validity of the master equation, obtained points of maximum are in agreement with the numerical simulation ($r\simeq0.2$ for the $+5$ and $r\simeq0.6$ for the $-3$ and $-7$ components). 

The physical origin of the maxima can also be seen directly from
Eq.~(\ref{app:eq:fmq}). The parameter $m$ measures the strength of the
pair-correlated mixing channel and decreases with decreasing $q$. At the same
time, $f$ increases because the normal fluorescence noise becomes weaker and
the probe is less broadened and less saturated by the source. Higher-order QWM
components contain different powers of $f$ and $m$, and therefore reach their
optimal values at different source bandwidths.

\newpage

% Bibliography generated with BibTeX (apsrev) and embedded for APS production.

\end{document}